\documentclass[aps,prl,12pt,floatfix,nofootinbib]{revtex4-1}
\usepackage{graphicx}
\usepackage{color}
\usepackage{multirow}
\usepackage{soul}
\usepackage{setspace}

\begin{document}
\title{ Dual origin of pairing in nuclei}
\author{A. Idini$^{a}$, G. Potel$^{b,c}$, F. Barranco$^{d}$,
 E. Vigezzi$^{e}$ and R.A. Broglia$^{f,g} $}
\affiliation{
$^a$  Institut f\"ur Kernphysik, Technische Universit\"at Darmstadt, Schlossgartenstrasse 2, 64289 Darmstadt, Germany\\
$^b$ National Superconducting Cyclotron Laboratory, Michigan State University, East Lansing, Michigan 48824, USA \\
$^c$ Lawrence Livermore National Laboratory, PO Box 808, L-414, Livermore, CA 94551, USA \\
$^d$ Departamento de Fisica Aplicada III, Escuela Superior de Ingenieros, Universidad de Sevilla, 
Camino de los Descubrimientos s/n,  
  41092 Sevilla, Spain.\\
$^e$ INFN, Sezione di Milano, Via Celoria 16, 20133 Milano, Italy.\\
$^f$ Diparitmento di Fisica, Universit\`a di Milano, Via Celoria 16, 20133 Milano, Italy. \\  
$^g$ The Niels Bohr Institute, University of Copenhagen, Blegdamsvej 17, Copenhagen, Denmark.\\  
}
\date{\today}

\begin{abstract}
%
An essentially "complete" description of the low-energy nuclear structure of the superfluid nucleus $^{120}$Sn and of its odd-$A$ neighbors is provided by the observations carried out with the help of Coulomb excitation and of one-- and of two-- particle transfer reactions, specific probes of vibrations, quasiparticle and pairing degrees of freedom respectively, and of their mutual couplings. These experimental findings are used to stringently test the predictions of a similarly "complete" description of $^{119,120,121}$Sn carried out in terms of elementary modes of excitation which, through their interweaving, melt together into effective fields, each displaying properties reflecting that of all others, there individuality resulting from the actual relative importance of each one. Its implementation is done by solving the Nambu-Gor'kov equations including, for the first time, all medium polarization effects resulting from the interweaving of quasiparticles, spin and surface vibrations, taking into account, within the framework of nuclear field theory (NFT), the variety of processes leading to self-energy, vertex and Pauli principle corrections, and to the induced pairing interaction. Theory provides an overall quantitative account of the experimental findings.
From these results one can, not only obtain strong circumstantial evidence for the inevitability for the dual origin of pairing in nuclei but also, extract information which can be used at profit to quantitatively disentangle the contributions to pairing correlations in general and to the pairing gap in particular, arising from the bare and from the induced pairing interactions.
\end{abstract}

\maketitle
\setstretch{2}

\begin{figure*}[ht]
   \centering		
 		\includegraphics[width=0.9\textwidth]{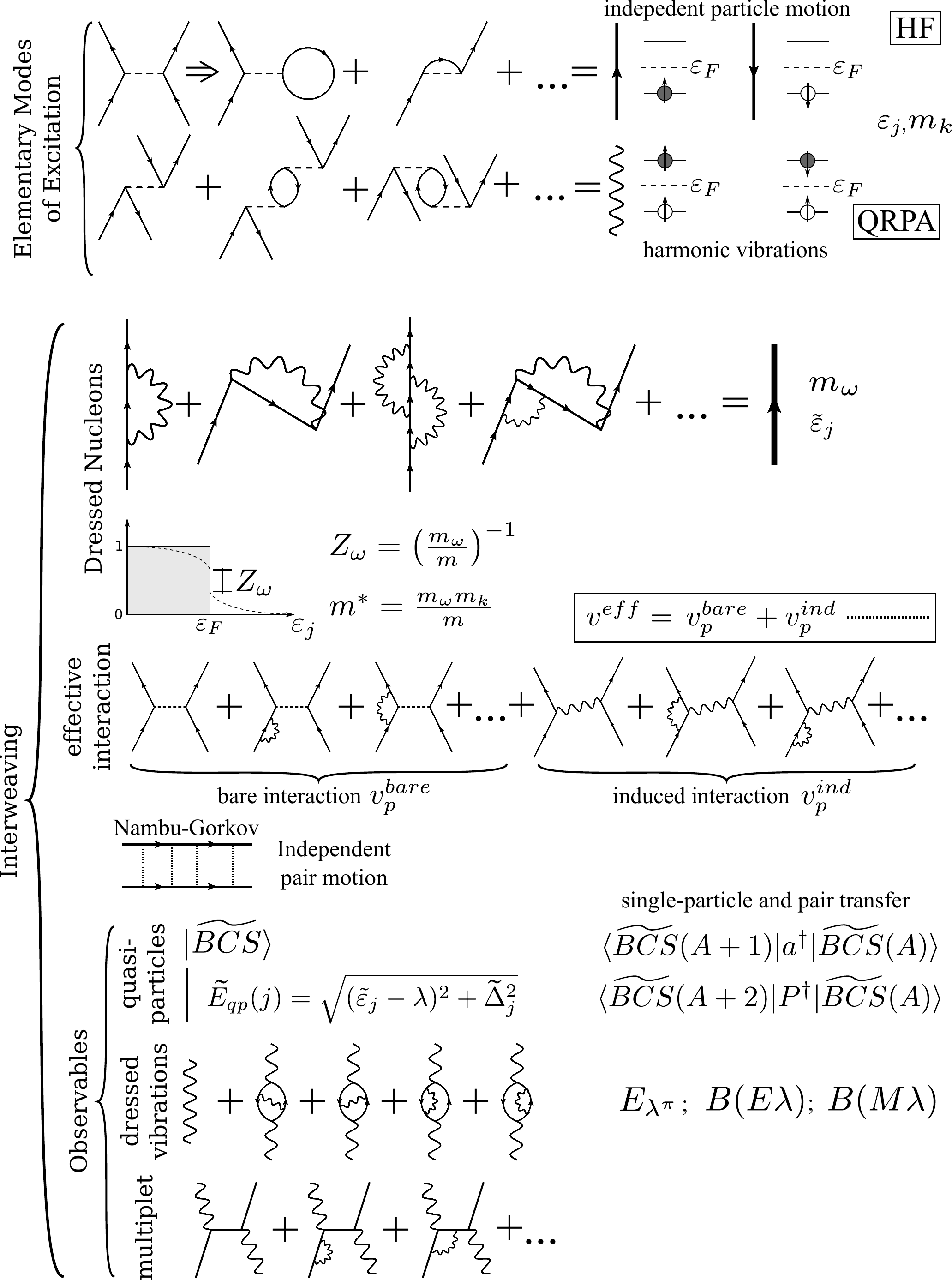}
 	\caption{Resum\'e of the strategy followed to calculate dressed elementary modes of excitation and induced pairing interaction in terms of NFT diagrams propagated by Nambu-Gorkov equations.}
 	\label{fig:NFT_Resume}
 \end{figure*}

\begin{figure}[htb]
  \centering		
		\includegraphics[width=0.70\textwidth]{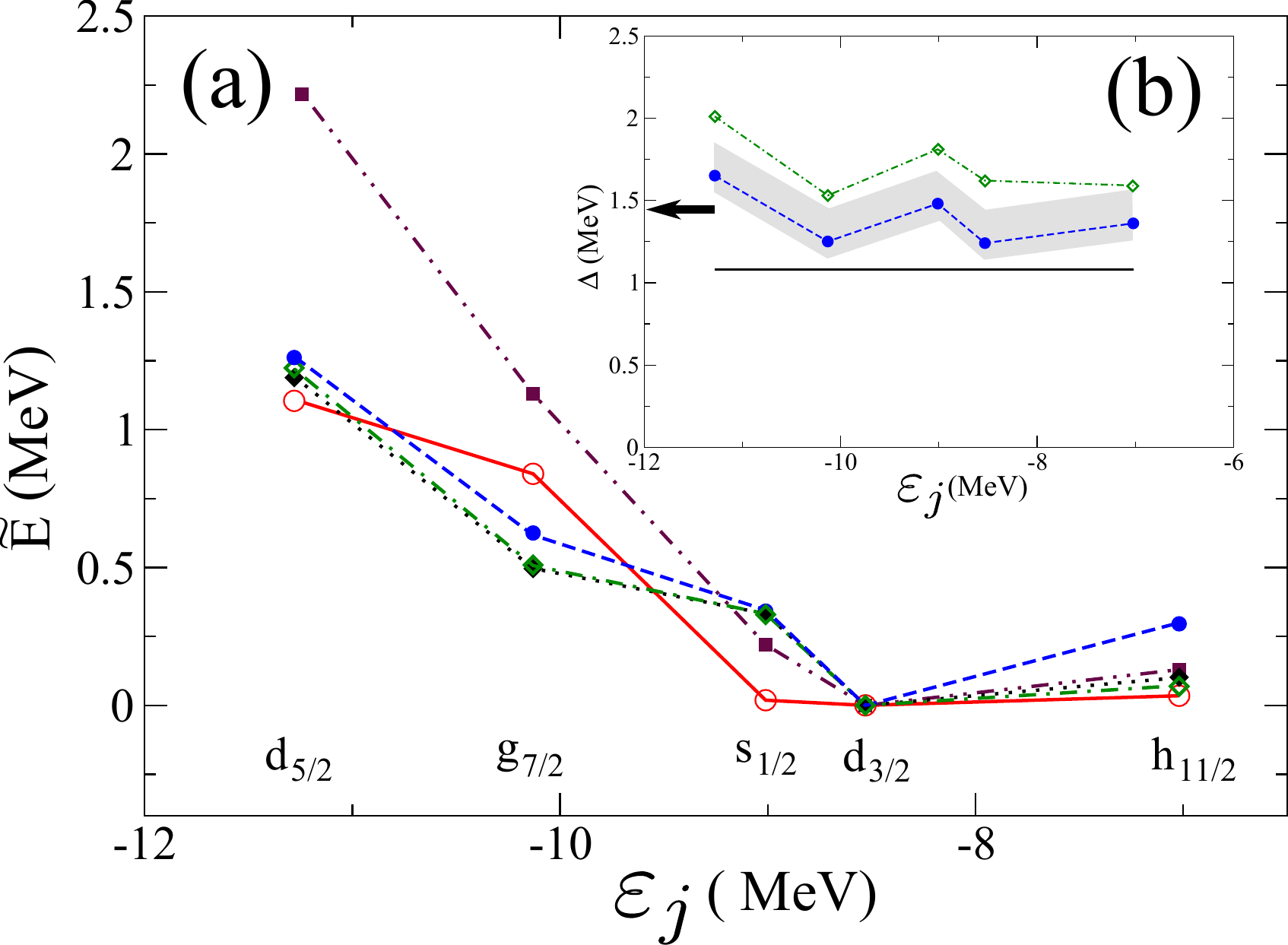}
	\caption{ (a) Lowest quasiparticle energies for the five valence  orbitals calculated for the nucleus $^{120}$Sn in comparison to the experimental energies (open circles), taken as  the average of the quasiparticle states observed in $^{121}$Sn and $^{119}$Sn. 
  The energies obtained including the bare interaction in the pairing channel with HF (SLy4) single--particle states, are shown by solid squares. 
The energies obtained by adding the coupling to surface modes are shown by solid diamonds. Adding vertex corrections leads to the results plotted in terms of open diamonds. The results obtained by adding the coupling to spin modes are displayed with solid dots.
	(b) Pairing gaps of the states lying around the Fermi energy.
        $\Delta^{BCS} ( = 1.08 $ MeV, horizontal straight line) was calculated with $v^{bare}_p$.  
	The effect of the coupling to surface modes is displayed in terms of open diamonds as in (a).
	The coupling to spin modes has also been included making use of the Landau parameters associated with the  SLy4 interaction
	(solid dots). The grey area shows the estimated errors of the calculations, due to variations associated with the choice of the (spin-spin) interaction and in the cutoff used to  calculate the coupling to spin modes. The empirical value (three-point formula) of the pairing gap is equal to 1.46 MeV (arrow on the vertical axis).}     
\label{fig:spectrum}
\end{figure}

Single-particle motion emerges from the same properties  which characterize  collective motion \cite{Bohr:75},  their interweaving leading 
to observables like quasiparticle states  and collective vibrations, and to renormalized interactions, in particular the dressed pairing interaction, 
 sum of the bare $v^{bare}_{p}$ interaction 
 and of the attraction resulting from the exchange  of vibrational modes between nucleons moving in time-reversal states lying close to the Fermi energy. 
 In this paper we calculate the pairing and the low--energy properties of $^{120}$Sn and of its neighboring odd-$A$ nuclei (see also \cite{Dobaczewski:13,Avogadro:13, Afdeenkov:13,Tarpanov:14} 
 and refs. therein). To this scope we diagonalize $v^{bare}_{p}$ in the BCS approximation making use of a basis composed  of nucleons  moving in the Hartree-Fock field resulting from the SLy4 interaction ($\varepsilon_{j},m_k$). We have fixed the strength of $v^{bare}_{p}$ by adopting the pairing gap calculated by Lesinski et al. \cite{Lesinski:12} (see also \cite{Holt:13}) fitting their results with a standard monopole-monopole pairing force of constant matrix elements \footnote{In this way one neglects the state dependence of the $NN+3N$ matrix elements which is of little consequence within the framework of the present paper (see e.g. \cite{Idini:12})}, the resulting value being $G=0.22 \textrm{ MeV}$. 
 We have calculated the vibrations of the system diagonalizing, in QRPA (two-quasiparticle basis) separable  multipole-multipole interactions  \cite{Bohr:75} (natural parity density modes with $\lambda^{\pi} = 2^+,3^-,4^+,5^-$)  and the spin-dependent part of the  SLy4 force  (spin modes of both natural and unnatural parity, $\lambda^{\pi} = 2^\pm,3^\mp,4^\pm,5^\mp$). 
   Taking into account  self-energy and renormalization processes according to the rules of NFT \cite{Bortignon:77,Bes:77,Bes:74,Broglia:76} (Fig. \ref{fig:NFT_Resume}), the dressed particle states
 (${\tilde  \varepsilon_{j}}, m_{\omega}, Z_{\omega}$) (cf. \cite{Mahaux:85}) and the induced  pairing interaction   $v_p^{ind}$ were calculated. 
 Adding $v^{ind}_p$ to $v^{bare}_{p}$,  the total effective pairing  interaction $v^{eff}_p$ was determined. 
 With these elements, the Nambu-Gor'kov (NG) equation \cite{Schrieffer:64,Idini:13,Idini:12,Soma:14} was solved selfconsistently using Green functions techniques, and the parameters characterizing the superfluid state determined. Within this framework, the quasiparticle energies $\tilde{E}_{\nu}$ and the state-dependent pairing gap $\tilde \Delta_{\nu}$ are related by the generalized gap equation
\begin{equation}
\tilde \Delta_{\nu}=  \tilde{\Delta}^{bare}_{\nu} + \tilde{\Delta}^{ind}_{\nu}= - Z_{\nu} \sum_{\nu'} \langle \nu' \bar \nu'| v^{bare}_p + v^{ind}_p| \nu \bar \nu \rangle  N_{\nu'}   
\frac{ \tilde \Delta_{\nu'}} {2  \tilde{E}_{\nu'}},
\label{delta_noi2}
\end{equation}
where $N_{\nu} = u^2_{\nu} +v^2_{\nu}$  and $Z_{\nu} = \left( 1 - \frac{\Sigma_{\nu}^{odd} }{E_{\nu} } \right)^{-1}$ 
($ \Sigma_{\nu}^{odd}$ being the odd part of the normal self-energy). Both the pairing gap and the lowest quasiparticle energies are displayed  in Fig. \ref{fig:spectrum} in comparison with the experimental findings. 
The contributions of $v^{bare}_p$ and $v^{ind}_p$ to $\tilde \Delta_{\nu}$ are about equal, density modes leading to attractive contributions which are partially canceled by that of spin modes\footnote{While we only display  the results obtained  with SLy4 to treat the spin modes,  the calculations were also carried out with other two effective forces, namely SkM$^*$ and SAMi. 
The errors attributed to the theoretical estimates  reflect the quantitative variation  of the cancellation effects.}, as expected from general transformation properties of the associated operators entering the particle vibration coupling vertices \cite{Bortignon:83}.
It is also noticed that $^{120}$Sn can be called  a Migdal "superconductor", in keeping with the fact  that vertex corrections are small throughout \cite{Migdal:58}.
 
\begin{figure}[htb]
  \centering		
		\includegraphics[width=0.90\textwidth]{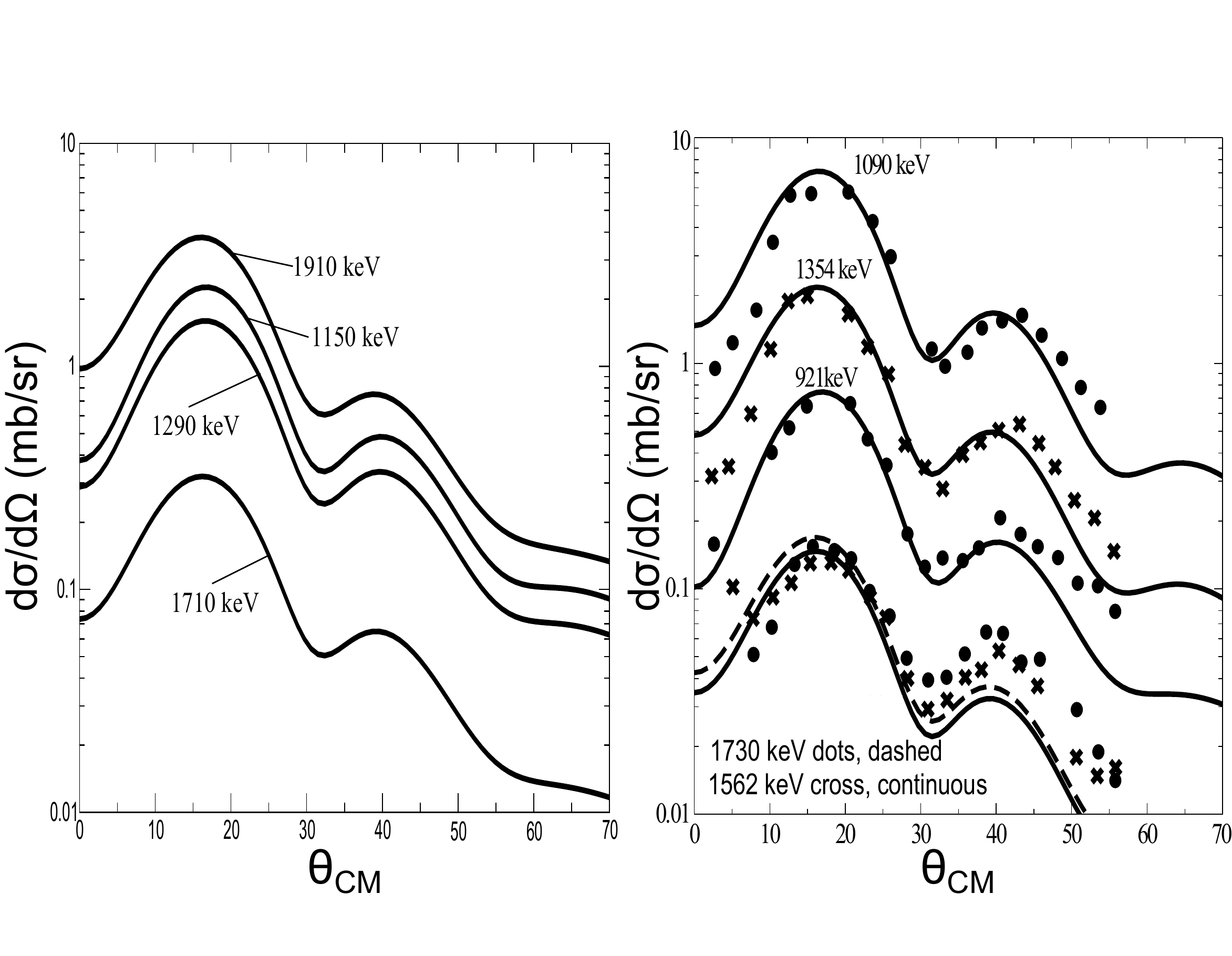}
	\caption{
	$^{120}$Sn$(p,d)^{119}$Sn$(5/2^+)$ absolute experimental cross sections \cite{Dickey:82} (dots), together with the DWBA fit carried out in the analysis of the data (right panel), in comparison with the finite range, full recoil DWBA calculations carried out with the help of state of the art optical potential and $v_{np}$ interaction (I. J. Thompson, private communication), making use of the NFT structure inputs as explained in the text.
	}
\label{fig:Reaction1}
\end{figure}

The strength functions associated  with the lowest  quasiparticle states provide an overall account of the experimental findings, in particular concerning the $u_{\nu}^2, v_{\nu}^2$ occupation numbers. Making use  of these quantities and of global optical parameters, the absolute one-particle  transfer cross sections were calculated. 
As an example, the results associated with the more trying and less well reproduced strength function, namely that of the $\tilde{E}_{d_{5/2}}$ quasiparticle state are displayed in Fig. \ref{fig:Reaction1} together with the experimental  data 
($^{120}$Sn(p,d)$^{119}$Sn) \cite{Dickey:82}. 
Four $d_{5/2}$ fragments are predicted by theory at low energy (\mbox{$< 2 \textrm{ MeV}$}) with a summed integrated cross section $\sum^4_{i = 1} \sigma(2^{\circ}-55^{\circ}) \approx 6.2 \textrm{ mb}$, while five are experimentally observed with $\sum^5_{i = 1} \sigma(2^{\circ}-55^{\circ}) \approx 8 \textrm{ mb} \pm 2 \textrm{ mb}$. 

\begin{figure}[htb]
  \centering		
  \includegraphics[width=0.95\textwidth]{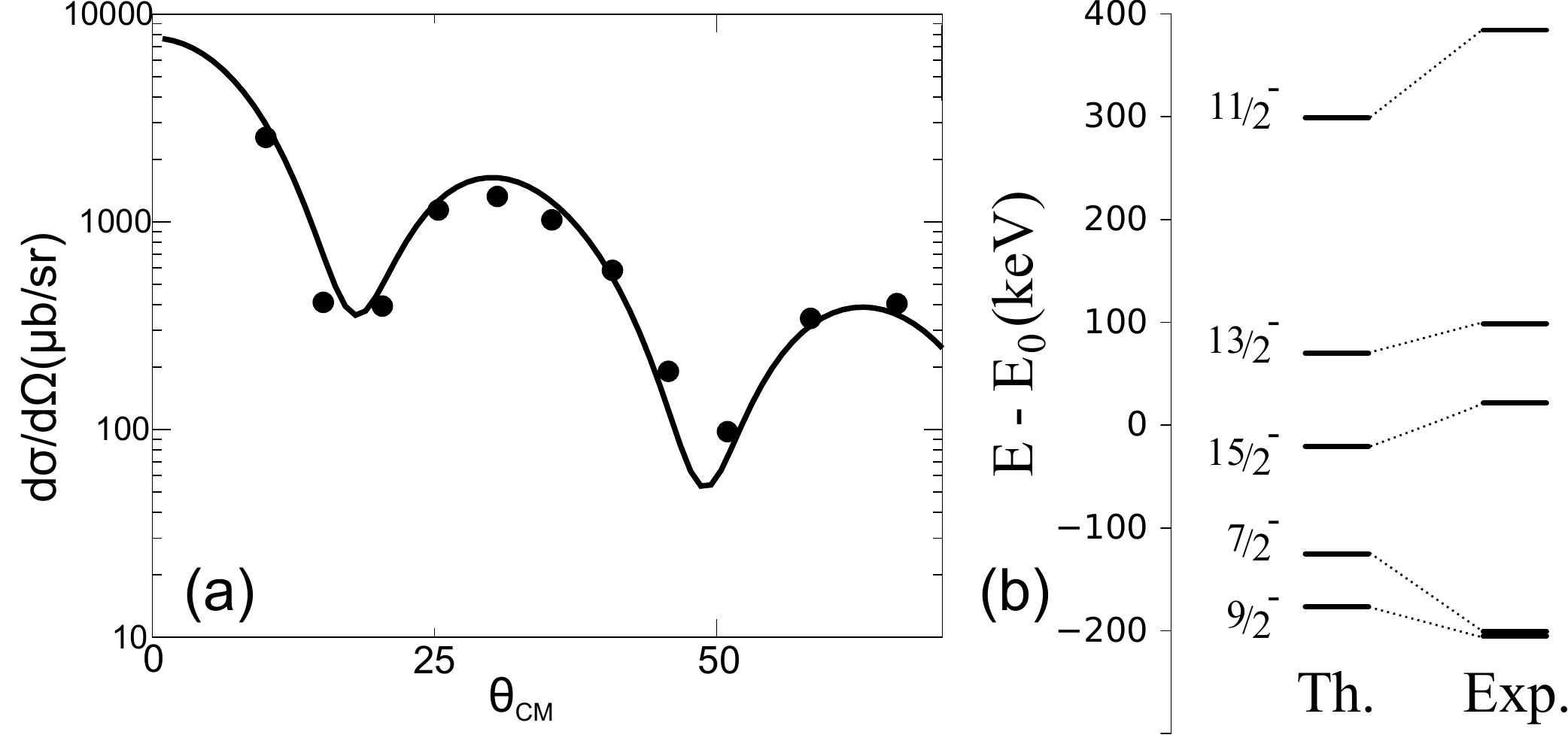}
  \caption{(a) Two particle transfer absolute cross section of $^{120}$Sn$(p,t)^{118}$Sn reaction second order DWBA calculations compared to experiment \cite{Guazzoni:08}. (b) Energies of the quintuplet of states arising from the coupling of the $h_{11/2}$ quasiparticle state and the lowest quadrupole vibration of $^{120}$Sn (i.e. $(h_{11/2} \otimes 2^{+})_{7/2^{-}-15/2^{-}}$) worked out using NFT diagrams. The experimental data is the average of $^{119}$Sn and $^{121}$Sn. The placement of the level $11/2^-$ member of the multiplet is from systematics \cite{BorelloLewin:75,Symochko:10,Ohya:10,Blachot:05}.}
\label{fig:Reaction2}
\label{fig:multiplet}
\end{figure}

 With the help of two-neutron spectroscopic amplitudes $B_{j_{\nu}} = (j_{\nu}+1/2)^{1/2} u_{j_{\nu}} v_{j_{\nu}}$ and of global optical parameters the absolute differential cross section 
 $^{120}$Sn(p,t)$^{118}$Sn(gs) was calculated and is plotted in Fig. \ref{fig:Reaction2}(a) in comparison with experimental data \cite{Guazzoni:08}. 
Theory ($\sigma(7.6^{\circ}-69.7^{\circ}) = 2360 \mu\textrm{b}$) reproduces the experimental integrated absolute cross section ($\sigma(7.6^{\circ}-69.7^{\circ}) = 2250 \pm 338 \mu\textrm{b}$) within experimental errors.
 
\begin{figure}[htb]
  \centering		
		\includegraphics[width=0.98\textwidth]{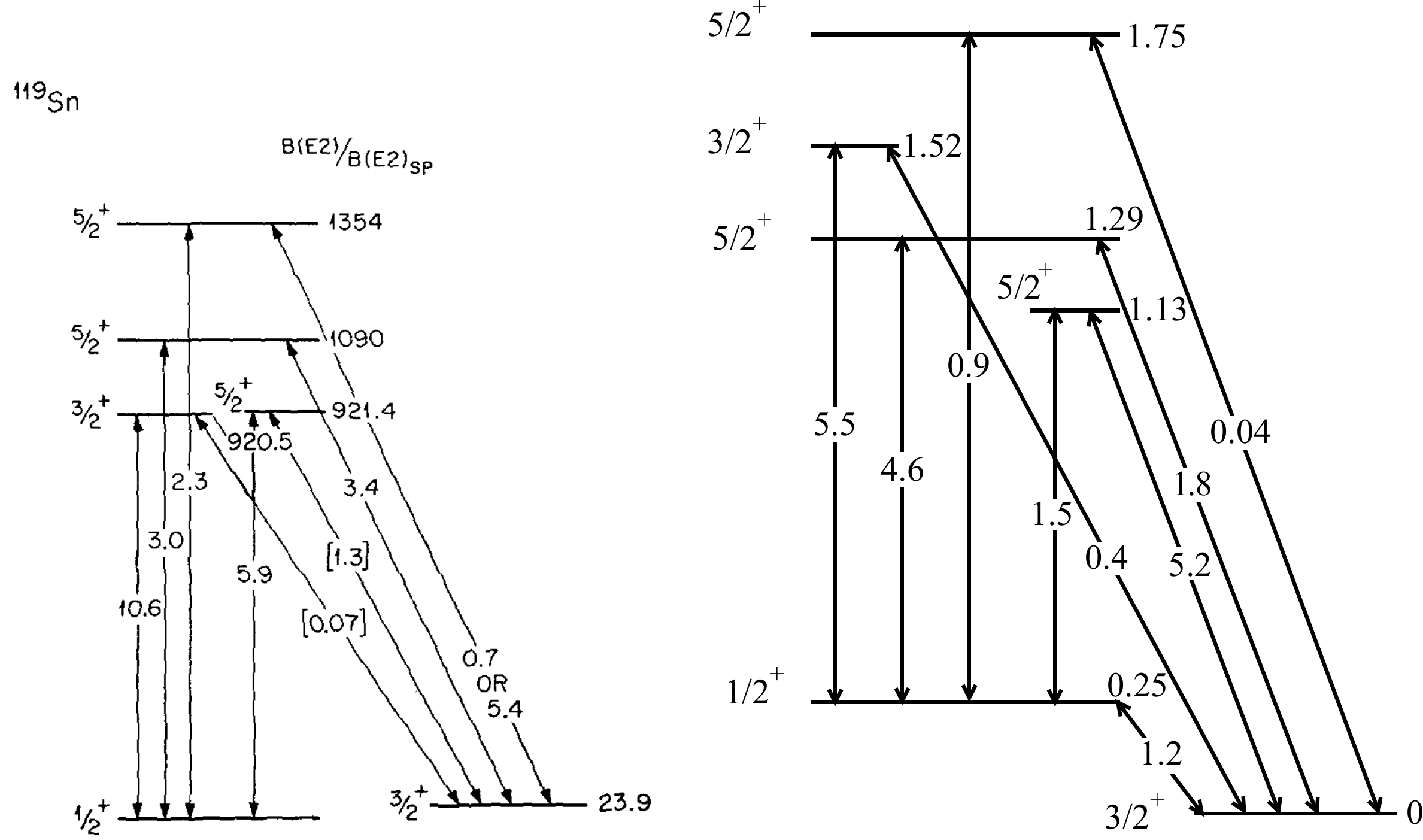}
	\caption{(left) Experimental level diagram of $^{119}$Sn and $B(E2 \downarrow)$ values in single-particle units, as
measured by Coulomb excitation \cite{Stelson:72}. (right) Theoretical spectrum.
 }     
\label{fig:ME2_decay}
\end{figure}

At the basis of the quantitative account of the experimental finding reported in Figs. 2, 3 and 4(a) one finds NFT renormalization processes in general, and induced pairing interaction in particular.
 One could argue that the evidences on which this statement is based are all related to virtual processes and thus amenable, in principle, to different interpretations. 
To answer this objection we have used the strength  of the coupling vertex $(h_{11/2} \otimes 2^+)_{ {15/2}^-  -  {7/2}^{-} }$ employed above, and calculated the  energy of the quintuplet of states arising from this coupling \cite{Bohr:75,Bortignon:77,Kuriyama:75}. 
Theory gives a good reproduction of the experimental data \cite{BorelloLewin:75,Symochko:10,Ohya:10,Blachot:05} (Fig. \ref{fig:multiplet}(b)) in a situation in which the basic states (quasiparticle and vibration) involved in the coupling are present asymptotically, the associated transfer processes making them real. This is also the case for the electromagnetic decay of the quasiparticle states. As seen from Fig. \ref{fig:ME2_decay}, theory also provides an overall quantitative picture of the Coulomb excitation measurements \cite{Stelson:72}.
 
 Summing up, structure properties, Coulomb excitation and one- and two- particle transfer reactions associated with  the superfluid nucleus $^{120}$Sn and with its neighboring odd-$A$ isotopes, have been calculated. 
 Both the bare pairing interaction as well as medium polarization effects resulting from the coupling of quasiparticles states with density and spin modes have been taken into account. The implementation of the program has been done using the NFT rules of structure and reactions \cite{Broglia:04a}, within the framework  of the Nambu-Gor'kov equations 
 ({\bf structure}), and of finite range, full recoil DWBA state of the art $(p,d)$ reaction theory, and of successive and simultaneous $(p,t)$ transfer reaction channels, corrected for  non-orthogonality contributions, within the framework of second order DWBA ({\bf reactions}) \cite{Potel:13,Potel:13b}. 
This is a first in the study of pairing in nuclei, 
theory providing a quantitative description of the nuclear response to a wide variety of external fields. 
Arguably, it also provides the first {\it bona fide} embodiment of the NFT renormalization program pioneered by Bohr, Mottelson, Bes and coworkers (cf. ref. \cite{Mottelson:76} and refs. therein).
Furthermore, the above results contain, through their essential ``completeness'', the basis to quantitatively assess the role played by $v^{bare}_p$ and $v^{ind}_p$ on the overall nuclear pairing correlations in general, and on the pairing gap in particular. In fact, changing the value of the different couplings contributing to $v^{ind}_p$, while boosting at the same time $v^{bare}_p$ so as to reproduce the empirical value of the (three-point formula) odd-even mass difference (solid arrow in Fig. \ref{fig:spectrum}(b)) would not alter the overall average value of $\tilde \Delta$, nor of the absolute two--particle transfer cross section (Fig. \ref{fig:Reaction2}(a)), but will change the splitting of the multiplet (Fig. \ref{fig:multiplet}(b)) and the value of the $B(E2)$ transition probabilities (Fig. \ref{fig:ME2_decay}), as well as the fragmentation of the $d_{5/2}$ quasiparticle state as revealed by the $^{120}$Sn$(p,d)^{119}$Sn$(5/2^+)$ cross sections (Fig. \ref{fig:Reaction1}).
Changing the different individual couplings in varied sequences and/or contemporaneities provides the basis for an operative protocol to disentangle the contribution of $v^{bare}_p$ and $v^{ind}_p$ to e.g. the pairing gap.

Within this context we conclude that pairing in typical superfluid nuclei lying along the stability valley like $^{119,120,121}$Sn has a dual origin, in which $v^{bare}_p$ and $v^{ind}_p$  play an essentially {\it on par} role.

RAB gratefully acknowledges extensive discussions with T. Duguet on the subjects which are at the basis of the present paper. FB and EV thank K. Matsuyanagi for illuminating discussions regarding the work of his group on quasiparticle--vibration coupling. AI is supported by the Helmholtz Association through the Nuclear Astrophysics Virtual Institute (VH-VI-417) and the Helmholtz International Center for FAIR within the framework of the LOEWE program launched by the state of Hesse.

\clearpage

\bibliographystyle{apsrev4-1} 
\bibliography{./nuclear}{}

\begin{thebibliography}{30}%
\makeatletter
\providecommand \@ifxundefined [1]{%
 \@ifx{#1\undefined}
}%
\providecommand \@ifnum [1]{%
 \ifnum #1\expandafter \@firstoftwo
 \else \expandafter \@secondoftwo
 \fi
}%
\providecommand \@ifx [1]{%
 \ifx #1\expandafter \@firstoftwo
 \else \expandafter \@secondoftwo
 \fi
}%
\providecommand \natexlab [1]{#1}%
\providecommand \enquote  [1]{``#1''}%
\providecommand \bibnamefont  [1]{#1}%
\providecommand \bibfnamefont [1]{#1}%
\providecommand \citenamefont [1]{#1}%
\providecommand \href@noop [0]{\@secondoftwo}%
\providecommand \href [0]{\begingroup \@sanitize@url \@href}%
\providecommand \@href[1]{\@@startlink{#1}\@@href}%
\providecommand \@@href[1]{\endgroup#1\@@endlink}%
\providecommand \@sanitize@url [0]{\catcode `\\12\catcode `\$12\catcode
  `\&12\catcode `\#12\catcode `\^12\catcode `\_12\catcode `\%12\relax}%
\providecommand \@@startlink[1]{}%
\providecommand \@@endlink[0]{}%
\providecommand \url  [0]{\begingroup\@sanitize@url \@url }%
\providecommand \@url [1]{\endgroup\@href {#1}{\urlprefix }}%
\providecommand \urlprefix  [0]{URL }%
\providecommand \Eprint [0]{\href }%
\providecommand \doibase [0]{http://dx.doi.org/}%
\providecommand \selectlanguage [0]{\@gobble}%
\providecommand \bibinfo  [0]{\@secondoftwo}%
\providecommand \bibfield  [0]{\@secondoftwo}%
\providecommand \translation [1]{[#1]}%
\providecommand \BibitemOpen [0]{}%
\providecommand \bibitemStop [0]{}%
\providecommand \bibitemNoStop [0]{.\EOS\space}%
\providecommand \EOS [0]{\spacefactor3000\relax}%
\providecommand \BibitemShut  [1]{\csname bibitem#1\endcsname}%
\let\auto@bib@innerbib\@empty
\bibitem [{\citenamefont {Bohr}\ and\ \citenamefont
  {Mottelson}(1975)}]{Bohr:75}%
  \BibitemOpen
  \bibfield  {author} {\bibinfo {author} {\bibfnamefont {A.}~\bibnamefont
  {Bohr}}\ and\ \bibinfo {author} {\bibfnamefont {B.~R.}\ \bibnamefont
  {Mottelson}},\ }\href@noop {} {\emph {\bibinfo {title} {Nuclear Structure,
  Vol.II}}}\ (\bibinfo  {publisher} {Benjamin},\ \bibinfo {address} {New
  York},\ \bibinfo {year} {1975})\BibitemShut {NoStop}%
\bibitem [{\citenamefont {Dobaczewski}\ and\ \citenamefont
  {Nazarewicz}(2013)}]{Dobaczewski:13}%
  \BibitemOpen
  \bibfield  {author} {\bibinfo {author} {\bibfnamefont {J.}~\bibnamefont
  {Dobaczewski}}\ and\ \bibinfo {author} {\bibfnamefont {W.}~\bibnamefont
  {Nazarewicz}},\ }in\ \href@noop {} {\emph {\bibinfo {booktitle} {Fifty Years
  of Nuclear {BCS}}}},\ \bibinfo {editor} {edited by\ \bibinfo {editor}
  {\bibfnamefont {R.~A.}\ \bibnamefont {Broglia}}\ and\ \bibinfo {editor}
  {\bibfnamefont {V.}~\bibnamefont {Zelevinsky}}}\ (\bibinfo  {publisher}
  {World Scientific Publishing Company},\ \bibinfo {year} {2013})\ p.~\bibinfo
  {pages} {40}\BibitemShut {NoStop}%
\bibitem [{\citenamefont {Avogadro}\ \emph {et~al.}(2013)\citenamefont
  {Avogadro}, \citenamefont {Barranco}, \citenamefont {Idini},\ and\
  \citenamefont {Vigezzi}}]{Avogadro:13}%
  \BibitemOpen
  \bibfield  {author} {\bibinfo {author} {\bibfnamefont {P.}~\bibnamefont
  {Avogadro}}, \bibinfo {author} {\bibfnamefont {F.}~\bibnamefont {Barranco}},
  \bibinfo {author} {\bibfnamefont {A.}~\bibnamefont {Idini}}, \ and\ \bibinfo
  {author} {\bibfnamefont {E.}~\bibnamefont {Vigezzi}},\ }in\ \href@noop {}
  {\emph {\bibinfo {booktitle} {Fifty Years of Nuclear {BCS}}}},\ \bibinfo
  {editor} {edited by\ \bibinfo {editor} {\bibfnamefont {R.~A.}\ \bibnamefont
  {Broglia}}\ and\ \bibinfo {editor} {\bibfnamefont {V.}~\bibnamefont
  {Zelevinsky}}}\ (\bibinfo  {publisher} {World Scientific, Singapore},\
  \bibinfo {year} {2013})\ p.\ \bibinfo {pages} {243}\BibitemShut {NoStop}%
\bibitem [{\citenamefont {Afdeenkov}\ and\ \citenamefont
  {Kamerdzhiev}(2013)}]{Afdeenkov:13}%
  \BibitemOpen
  \bibfield  {author} {\bibinfo {author} {\bibfnamefont {A.}~\bibnamefont
  {Afdeenkov}}\ and\ \bibinfo {author} {\bibfnamefont {S.}~\bibnamefont
  {Kamerdzhiev}},\ }in\ \href@noop {} {\emph {\bibinfo {booktitle} {Fifty Years
  of Nuclear {BCS}}}},\ \bibinfo {editor} {edited by\ \bibinfo {editor}
  {\bibfnamefont {R.~A.}\ \bibnamefont {Broglia}}\ and\ \bibinfo {editor}
  {\bibfnamefont {V.}~\bibnamefont {Zelevinsky}}}\ (\bibinfo  {publisher}
  {World Scientific Publishing, Singapore},\ \bibinfo {year} {2013})\ p.\
  \bibinfo {pages} {274}\BibitemShut {NoStop}%
\bibitem [{\citenamefont {Tarpanov}\ \emph {et~al.}(2014)\citenamefont
  {Tarpanov}, \citenamefont {Toivanen}, \citenamefont {Dobaczewski},\ and\
  \citenamefont {Carlsson}}]{Tarpanov:14}%
  \BibitemOpen
  \bibfield  {author} {\bibinfo {author} {\bibfnamefont {D.}~\bibnamefont
  {Tarpanov}}, \bibinfo {author} {\bibfnamefont {J.}~\bibnamefont {Toivanen}},
  \bibinfo {author} {\bibfnamefont {J.}~\bibnamefont {Dobaczewski}}, \ and\
  \bibinfo {author} {\bibfnamefont {B.~G.}\ \bibnamefont {Carlsson}},\ }\href
  {\doibase 10.1103/PhysRevC.89.014307} {\bibfield  {journal} {\bibinfo
  {journal} {Phys. Rev. C}\ }\textbf {\bibinfo {volume} {89}},\ \bibinfo
  {pages} {014307} (\bibinfo {year} {2014})}\BibitemShut {NoStop}%
\bibitem [{\citenamefont {Lesinski}\ \emph {et~al.}(2012)\citenamefont
  {Lesinski}, \citenamefont {Hebeler}, \citenamefont {Duguet},\ and\
  \citenamefont {Schwenk}}]{Lesinski:12}%
  \BibitemOpen
  \bibfield  {author} {\bibinfo {author} {\bibfnamefont {T.}~\bibnamefont
  {Lesinski}}, \bibinfo {author} {\bibfnamefont {K.}~\bibnamefont {Hebeler}},
  \bibinfo {author} {\bibfnamefont {T.}~\bibnamefont {Duguet}}, \ and\ \bibinfo
  {author} {\bibfnamefont {A.}~\bibnamefont {Schwenk}},\ }\href {\doibase
  doi:10.1088/0954-3899/39/1/015108} {\bibfield  {journal} {\bibinfo  {journal}
  {J. Phys. G}\ }\textbf {\bibinfo {volume} {39}},\ \bibinfo {pages} {15108}
  (\bibinfo {year} {2012})}\BibitemShut {NoStop}%
\bibitem [{\citenamefont {Holt}\ \emph {et~al.}(2013)\citenamefont {Holt},
  \citenamefont {Men\'{e}ndez},\ and\ \citenamefont {Schwenk}}]{Holt:13}%
  \BibitemOpen
  \bibfield  {author} {\bibinfo {author} {\bibfnamefont {J.~D.}\ \bibnamefont
  {Holt}}, \bibinfo {author} {\bibfnamefont {J.}~\bibnamefont {Men\'{e}ndez}},
  \ and\ \bibinfo {author} {\bibfnamefont {A.}~\bibnamefont {Schwenk}},\ }\href
  {http://stacks.iop.org/0954-3899/40/i=7/a=075105} {\bibfield  {journal}
  {\bibinfo  {journal} {J. Phys. G}\ }\textbf {\bibinfo {volume} {40}},\
  \bibinfo {pages} {075105} (\bibinfo {year} {2013})}\BibitemShut {NoStop}%
\bibitem [{\citenamefont {Idini}\ \emph {et~al.}(2012)\citenamefont {Idini},
  \citenamefont {Barranco},\ and\ \citenamefont {Vigezzi}}]{Idini:12}%
  \BibitemOpen
  \bibfield  {author} {\bibinfo {author} {\bibfnamefont {A.}~\bibnamefont
  {Idini}}, \bibinfo {author} {\bibfnamefont {F.}~\bibnamefont {Barranco}}, \
  and\ \bibinfo {author} {\bibfnamefont {E.}~\bibnamefont {Vigezzi}},\ }\href
  {\doibase 10.1103/PhysRevC.85.014331} {\bibfield  {journal} {\bibinfo
  {journal} {Phys. Rev. C}\ }\textbf {\bibinfo {volume} {85}},\ \bibinfo
  {pages} {014331} (\bibinfo {year} {2012})}\BibitemShut {NoStop}%
\bibitem [{\citenamefont {Bortignon}\ \emph {et~al.}(1977)\citenamefont
  {Bortignon}, \citenamefont {Broglia}, \citenamefont {B{\`{e}}s},\ and\
  \citenamefont {Liotta}}]{Bortignon:77}%
  \BibitemOpen
  \bibfield  {author} {\bibinfo {author} {\bibfnamefont {P.~F.}\ \bibnamefont
  {Bortignon}}, \bibinfo {author} {\bibfnamefont {R.~A.}\ \bibnamefont
  {Broglia}}, \bibinfo {author} {\bibfnamefont {D.~R.}\ \bibnamefont
  {B{\`{e}}s}}, \ and\ \bibinfo {author} {\bibfnamefont {R.}~\bibnamefont
  {Liotta}},\ }\href@noop {} {\bibfield  {journal} {\bibinfo  {journal} {Phys.
  Rep.}\ }\textbf {\bibinfo {volume} {30}},\ \bibinfo {pages} {305} (\bibinfo
  {year} {1977})}\BibitemShut {NoStop}%
\bibitem [{\citenamefont {B{\`{e}}s}\ and\ \citenamefont
  {Broglia}(1977)}]{Bes:77}%
  \BibitemOpen
  \bibfield  {author} {\bibinfo {author} {\bibfnamefont {D.~R.}\ \bibnamefont
  {B{\`{e}}s}}\ and\ \bibinfo {author} {\bibfnamefont {R.~A.}\ \bibnamefont
  {Broglia}},\ }in\ \href@noop {} {\emph {\bibinfo {booktitle} {International
  School of Physics ``Enrico Fermi'' Course LXIX, Elementary Modes of
  Excitation in Nuclei}}},\ \bibinfo {editor} {edited by\ \bibinfo {editor}
  {\bibfnamefont {A.}~\bibnamefont {Bohr}}\ and\ \bibinfo {editor}
  {\bibfnamefont {R.~A.}\ \bibnamefont {Broglia}}}\ (\bibinfo  {publisher}
  {North Holland},\ \bibinfo {address} {Amsterdam},\ \bibinfo {year} {1977})\
  p.~\bibinfo {pages} {55}\BibitemShut {NoStop}%
\bibitem [{\citenamefont {Bes}\ \emph {et~al.}(1974)\citenamefont {Bes},
  \citenamefont {Dussel}, \citenamefont {Broglia}, \citenamefont {Liotta},\
  and\ \citenamefont {Mottelson}}]{Bes:74}%
  \BibitemOpen
  \bibfield  {author} {\bibinfo {author} {\bibfnamefont {D.~R.}\ \bibnamefont
  {Bes}}, \bibinfo {author} {\bibfnamefont {G.~G.}\ \bibnamefont {Dussel}},
  \bibinfo {author} {\bibfnamefont {R.~A.}\ \bibnamefont {Broglia}}, \bibinfo
  {author} {\bibfnamefont {R.}~\bibnamefont {Liotta}}, \ and\ \bibinfo {author}
  {\bibfnamefont {B.~R.}\ \bibnamefont {Mottelson}},\ }\href@noop {} {\bibfield
   {journal} {\bibinfo  {journal} {Phys. Lett. B}\ }\textbf {\bibinfo {volume}
  {52}},\ \bibinfo {pages} {253} (\bibinfo {year} {1974})}\BibitemShut
  {NoStop}%
\bibitem [{\citenamefont {Broglia}\ \emph {et~al.}(1976)\citenamefont
  {Broglia}, \citenamefont {Mottelson}, \citenamefont {B{\`{e}}s},
  \citenamefont {Liotta},\ and\ \citenamefont {Sof{\'{\i}}a}}]{Broglia:76}%
  \BibitemOpen
  \bibfield  {author} {\bibinfo {author} {\bibfnamefont {R.~A.}\ \bibnamefont
  {Broglia}}, \bibinfo {author} {\bibfnamefont {B.~R.}\ \bibnamefont
  {Mottelson}}, \bibinfo {author} {\bibfnamefont {D.~R.}\ \bibnamefont
  {B{\`{e}}s}}, \bibinfo {author} {\bibfnamefont {R.}~\bibnamefont {Liotta}}, \
  and\ \bibinfo {author} {\bibfnamefont {H.~M.}\ \bibnamefont {Sof{\'{\i}}a}},\
  }\href@noop {} {\bibfield  {journal} {\bibinfo  {journal} {Physics Letters
  B}\ }\textbf {\bibinfo {volume} {64}},\ \bibinfo {pages} {29} (\bibinfo
  {year} {1976})}\BibitemShut {NoStop}%
\bibitem [{\citenamefont {Mahaux}\ \emph {et~al.}(1985)\citenamefont {Mahaux},
  \citenamefont {Bortignon}, \citenamefont {Broglia},\ and\ \citenamefont
  {Dasso}}]{Mahaux:85}%
  \BibitemOpen
  \bibfield  {author} {\bibinfo {author} {\bibfnamefont {C.}~\bibnamefont
  {Mahaux}}, \bibinfo {author} {\bibfnamefont {P.~F.}\ \bibnamefont
  {Bortignon}}, \bibinfo {author} {\bibfnamefont {R.~A.}\ \bibnamefont
  {Broglia}}, \ and\ \bibinfo {author} {\bibfnamefont {C.~H.}\ \bibnamefont
  {Dasso}},\ }\href@noop {} {\bibfield  {journal} {\bibinfo  {journal} {Phys.
  Rep.}\ }\textbf {\bibinfo {volume} {120}},\ \bibinfo {pages} {1} (\bibinfo
  {year} {1985})}\BibitemShut {NoStop}%
\bibitem [{\citenamefont {Schrieffer}(1964)}]{Schrieffer:64}%
  \BibitemOpen
  \bibfield  {author} {\bibinfo {author} {\bibfnamefont {J.}~\bibnamefont
  {Schrieffer}},\ }\href@noop {} {\emph {\bibinfo {title}
  {Superconductivity}}}\ (\bibinfo  {publisher} {Benjamin},\ \bibinfo {address}
  {New York},\ \bibinfo {year} {1964})\BibitemShut {NoStop}%
\bibitem [{\citenamefont {Idini}(2013)}]{Idini:13}%
  \BibitemOpen
  \bibfield  {author} {\bibinfo {author} {\bibfnamefont {A.}~\bibnamefont
  {Idini}},\ }\emph {\bibinfo {title} {Renormalization effects in nuclei}},\
  \href@noop {} {Ph.D. thesis},\ \bibinfo  {school} {University of Milan}
  (\bibinfo {year} {2013}),\ \bibinfo {note}
  {http://air.unimi.it/handle/2434/216315}\BibitemShut {NoStop}%
\bibitem [{\citenamefont {Som\`a}\ \emph {et~al.}(2014)\citenamefont {Som\`a},
  \citenamefont {Barbieri},\ and\ \citenamefont {Duguet}}]{Soma:14}%
  \BibitemOpen
  \bibfield  {author} {\bibinfo {author} {\bibfnamefont {V.}~\bibnamefont
  {Som\`a}}, \bibinfo {author} {\bibfnamefont {C.}~\bibnamefont {Barbieri}}, \
  and\ \bibinfo {author} {\bibfnamefont {T.}~\bibnamefont {Duguet}},\ }\href
  {\doibase 10.1103/PhysRevC.89.024323} {\bibfield  {journal} {\bibinfo
  {journal} {Phys. Rev. C}\ }\textbf {\bibinfo {volume} {89}},\ \bibinfo
  {pages} {024323} (\bibinfo {year} {2014})}\BibitemShut {NoStop}%
\bibitem [{\citenamefont {Bortignon}\ \emph {et~al.}(1983)\citenamefont
  {Bortignon}, \citenamefont {Broglia},\ and\ \citenamefont
  {Dasso}}]{Bortignon:83}%
  \BibitemOpen
  \bibfield  {author} {\bibinfo {author} {\bibfnamefont {P.~F.}\ \bibnamefont
  {Bortignon}}, \bibinfo {author} {\bibfnamefont {R.~A.}\ \bibnamefont
  {Broglia}}, \ and\ \bibinfo {author} {\bibfnamefont {C.}~\bibnamefont
  {Dasso}},\ }\href {\doibase http://dx.doi.org/10.1016/0375-9474(83)90484-0}
  {\bibfield  {journal} {\bibinfo  {journal} {Nuclear Physics A}\ }\textbf
  {\bibinfo {volume} {398}},\ \bibinfo {pages} {221} (\bibinfo {year}
  {1983})}\BibitemShut {NoStop}%
\bibitem [{\citenamefont {Migdal}(1958)}]{Migdal:58}%
  \BibitemOpen
  \bibfield  {author} {\bibinfo {author} {\bibfnamefont {A.~B.}\ \bibnamefont
  {Migdal}},\ }\href@noop {} {\bibfield  {journal} {\bibinfo  {journal} {Sov.
  Phys. JETP}\ }\textbf {\bibinfo {volume} {7}},\ \bibinfo {pages} {996}
  (\bibinfo {year} {1958})}\BibitemShut {NoStop}%
\bibitem [{\citenamefont {Dickey}\ \emph {et~al.}(1982)\citenamefont {Dickey},
  \citenamefont {Kraushaar}, \citenamefont {Ristinen},\ and\ \citenamefont
  {Rumore}}]{Dickey:82}%
  \BibitemOpen
  \bibfield  {author} {\bibinfo {author} {\bibfnamefont {S.}~\bibnamefont
  {Dickey}}, \bibinfo {author} {\bibfnamefont {J.}~\bibnamefont {Kraushaar}},
  \bibinfo {author} {\bibfnamefont {R.}~\bibnamefont {Ristinen}}, \ and\
  \bibinfo {author} {\bibfnamefont {M.}~\bibnamefont {Rumore}},\ }\href
  {\doibase 10.1016/0375-9474(82)90325-6} {\bibfield  {journal} {\bibinfo
  {journal} {Nuclear Physics A}\ }\textbf {\bibinfo {volume} {377}},\ \bibinfo
  {pages} {137} (\bibinfo {year} {1982})}\BibitemShut {NoStop}%
\bibitem [{\citenamefont {Guazzoni}\ \emph {et~al.}(2008)\citenamefont
  {Guazzoni}, \citenamefont {Zetta}, \citenamefont {Covello}, \citenamefont
  {Gargano}, \citenamefont {Bayman}, \citenamefont {Faestermann}, \citenamefont
  {Graw}, \citenamefont {Hertenberger}, \citenamefont {Wirth},\ and\
  \citenamefont {Jaskola}}]{Guazzoni:08}%
  \BibitemOpen
  \bibfield  {author} {\bibinfo {author} {\bibfnamefont {P.}~\bibnamefont
  {Guazzoni}}, \bibinfo {author} {\bibfnamefont {L.}~\bibnamefont {Zetta}},
  \bibinfo {author} {\bibfnamefont {A.}~\bibnamefont {Covello}}, \bibinfo
  {author} {\bibfnamefont {A.}~\bibnamefont {Gargano}}, \bibinfo {author}
  {\bibfnamefont {B.~F.}\ \bibnamefont {Bayman}}, \bibinfo {author}
  {\bibfnamefont {T.}~\bibnamefont {Faestermann}}, \bibinfo {author}
  {\bibfnamefont {G.}~\bibnamefont {Graw}}, \bibinfo {author} {\bibfnamefont
  {R.}~\bibnamefont {Hertenberger}}, \bibinfo {author} {\bibfnamefont {H.-F.}\
  \bibnamefont {Wirth}}, \ and\ \bibinfo {author} {\bibfnamefont
  {M.}~\bibnamefont {Jaskola}},\ }\href {\doibase 10.1103/PhysRevC.78.064608}
  {\bibfield  {journal} {\bibinfo  {journal} {Phys. Rev. C}\ }\textbf {\bibinfo
  {volume} {78}},\ \bibinfo {pages} {064608} (\bibinfo {year}
  {2008})}\BibitemShut {NoStop}%
\bibitem [{\citenamefont {Kuriyama}\ \emph {et~al.}(1975)\citenamefont
  {Kuriyama}, \citenamefont {Marumori},\ and\ \citenamefont
  {Matsuyanagi}}]{Kuriyama:75}%
  \BibitemOpen
  \bibfield  {author} {\bibinfo {author} {\bibfnamefont {A.}~\bibnamefont
  {Kuriyama}}, \bibinfo {author} {\bibfnamefont {T.}~\bibnamefont {Marumori}},
  \ and\ \bibinfo {author} {\bibfnamefont {K.}~\bibnamefont {Matsuyanagi}},\
  }\href {\doibase 10.1143/PTPS.58.53} {\bibfield  {journal} {\bibinfo
  {journal} {Progress of Theoretical Physics Supplement}\ }\textbf {\bibinfo
  {volume} {58}},\ \bibinfo {pages} {53} (\bibinfo {year} {1975})}\BibitemShut
  {NoStop}%
\bibitem [{\citenamefont {Borello-Lewin}\ \emph {et~al.}(1975)\citenamefont
  {Borello-Lewin}, \citenamefont {Orsini}, \citenamefont {Dietzsch},\ and\
  \citenamefont {Hamburger}}]{BorelloLewin:75}%
  \BibitemOpen
  \bibfield  {author} {\bibinfo {author} {\bibfnamefont {T.}~\bibnamefont
  {Borello-Lewin}}, \bibinfo {author} {\bibfnamefont {C.}~\bibnamefont
  {Orsini}}, \bibinfo {author} {\bibfnamefont {O.}~\bibnamefont {Dietzsch}}, \
  and\ \bibinfo {author} {\bibfnamefont {E.}~\bibnamefont {Hamburger}},\ }\href
  {\doibase http://dx.doi.org/10.1016/0375-9474(75)90188-8} {\bibfield
  {journal} {\bibinfo  {journal} {Nuclear Physics A}\ }\textbf {\bibinfo
  {volume} {249}},\ \bibinfo {pages} {284} (\bibinfo {year}
  {1975})}\BibitemShut {NoStop}%
\bibitem [{\citenamefont {Symochko}\ \emph {et~al.}(2010)\citenamefont
  {Symochko}, \citenamefont {Browne},\ and\ \citenamefont
  {Tuli}}]{Symochko:10}%
  \BibitemOpen
  \bibfield  {author} {\bibinfo {author} {\bibfnamefont {D.~M.}\ \bibnamefont
  {Symochko}}, \bibinfo {author} {\bibfnamefont {E.}~\bibnamefont {Browne}}, \
  and\ \bibinfo {author} {\bibfnamefont {J.~K.}\ \bibnamefont {Tuli}},\
  }\href@noop {} {\bibfield  {journal} {\bibinfo  {journal} {Nuclear Data
  Sheets}\ }\textbf {\bibinfo {volume} {110}},\ \bibinfo {pages} {2945}
  (\bibinfo {year} {2010})}\BibitemShut {NoStop}%
\bibitem [{\citenamefont {Ohya}(2010)}]{Ohya:10}%
  \BibitemOpen
  \bibfield  {author} {\bibinfo {author} {\bibfnamefont {S.}~\bibnamefont
  {Ohya}},\ }\href@noop {} {\bibfield  {journal} {\bibinfo  {journal} {Nuclear
  Data Sheets}\ }\textbf {\bibinfo {volume} {111}},\ \bibinfo {pages} {1619}
  (\bibinfo {year} {2010})}\BibitemShut {NoStop}%
\bibitem [{\citenamefont {Blachot}(2005)}]{Blachot:05}%
  \BibitemOpen
  \bibfield  {author} {\bibinfo {author} {\bibfnamefont {J.}~\bibnamefont
  {Blachot}},\ }\href@noop {} {\bibfield  {journal} {\bibinfo  {journal}
  {Nuclear Data Sheets}\ }\textbf {\bibinfo {volume} {104}},\ \bibinfo {pages}
  {967} (\bibinfo {year} {2005})}\BibitemShut {NoStop}%
\bibitem [{\citenamefont {Stelson}\ \emph {et~al.}(1972)\citenamefont
  {Stelson}, \citenamefont {Milner}, \citenamefont {McGowan}, \citenamefont
  {Robinson},\ and\ \citenamefont {Raman}}]{Stelson:72}%
  \BibitemOpen
  \bibfield  {author} {\bibinfo {author} {\bibfnamefont {P.}~\bibnamefont
  {Stelson}}, \bibinfo {author} {\bibfnamefont {W.}~\bibnamefont {Milner}},
  \bibinfo {author} {\bibfnamefont {F.}~\bibnamefont {McGowan}}, \bibinfo
  {author} {\bibfnamefont {R.}~\bibnamefont {Robinson}}, \ and\ \bibinfo
  {author} {\bibfnamefont {S.}~\bibnamefont {Raman}},\ }\href {\doibase
  http://dx.doi.org/10.1016/0375-9474(72)90137-6} {\bibfield  {journal}
  {\bibinfo  {journal} {Nuclear Physics A}\ }\textbf {\bibinfo {volume}
  {190}},\ \bibinfo {pages} {197} (\bibinfo {year} {1972})}\BibitemShut
  {NoStop}%
\bibitem [{\citenamefont {Broglia}\ and\ \citenamefont
  {Winther}(2004)}]{Broglia:04a}%
  \BibitemOpen
  \bibfield  {author} {\bibinfo {author} {\bibfnamefont {R.~A.}\ \bibnamefont
  {Broglia}}\ and\ \bibinfo {author} {\bibfnamefont {A.}~\bibnamefont
  {Winther}},\ }\href@noop {} {\emph {\bibinfo {title} {Heavy Ion Reactions}}}\
  (\bibinfo  {publisher} {Westview Press},\ \bibinfo {address} {Cambridge,
  MA.},\ \bibinfo {year} {2004})\BibitemShut {NoStop}%
\bibitem [{\citenamefont {Potel}\ \emph
  {et~al.}(2013{\natexlab{a}})\citenamefont {Potel}, \citenamefont {Idini},
  \citenamefont {Barranco}, \citenamefont {Vigezzi},\ and\ \citenamefont
  {Broglia}}]{Potel:13}%
  \BibitemOpen
  \bibfield  {author} {\bibinfo {author} {\bibfnamefont {G.}~\bibnamefont
  {Potel}}, \bibinfo {author} {\bibfnamefont {A.}~\bibnamefont {Idini}},
  \bibinfo {author} {\bibfnamefont {F.}~\bibnamefont {Barranco}}, \bibinfo
  {author} {\bibfnamefont {E.}~\bibnamefont {Vigezzi}}, \ and\ \bibinfo
  {author} {\bibfnamefont {R.~A.}\ \bibnamefont {Broglia}},\ }\href
  {http://stacks.iop.org/0034-4885/76/i=10/a=106301} {\bibfield  {journal}
  {\bibinfo  {journal} {Reports on Progress in Physics}\ }\textbf {\bibinfo
  {volume} {76}},\ \bibinfo {pages} {106301} (\bibinfo {year}
  {2013}{\natexlab{a}})}\BibitemShut {NoStop}%
\bibitem [{\citenamefont {Potel}\ \emph
  {et~al.}(2013{\natexlab{b}})\citenamefont {Potel}, \citenamefont {Idini},
  \citenamefont {Barranco}, \citenamefont {Vigezzi},\ and\ \citenamefont
  {Broglia}}]{Potel:13b}%
  \BibitemOpen
  \bibfield  {author} {\bibinfo {author} {\bibfnamefont {G.}~\bibnamefont
  {Potel}}, \bibinfo {author} {\bibfnamefont {A.}~\bibnamefont {Idini}},
  \bibinfo {author} {\bibfnamefont {F.}~\bibnamefont {Barranco}}, \bibinfo
  {author} {\bibfnamefont {E.}~\bibnamefont {Vigezzi}}, \ and\ \bibinfo
  {author} {\bibfnamefont {R.~A.}\ \bibnamefont {Broglia}},\ }\href {\doibase
  10.1103/PhysRevC.87.054321} {\bibfield  {journal} {\bibinfo  {journal} {Phys.
  Rev. C}\ }\textbf {\bibinfo {volume} {87}},\ \bibinfo {pages} {054321}
  (\bibinfo {year} {2013}{\natexlab{b}})}\BibitemShut {NoStop}%
\bibitem [{\citenamefont {Mottelson}(1976)}]{Mottelson:76}%
  \BibitemOpen
  \bibfield  {author} {\bibinfo {author} {\bibfnamefont {B.~R.}\ \bibnamefont
  {Mottelson}},\ }\href@noop {} {\emph {\bibinfo {title} {Elementary Modes of
  Excitation in Nuclei, Le Prix Nobel en 1975}}},\ edited by\ \bibinfo {editor}
  {\bibfnamefont {W.}~\bibnamefont {Odelberg}}\ (\bibinfo  {publisher}
  {Imprimerie Royale Norstedts Tryckeri},\ \bibinfo {address} {Stockholm},\
  \bibinfo {year} {1976})\ \bibinfo {note} {p. 80}\BibitemShut {NoStop}%
\end{thebibliography}%

 \end{document}